\documentclass[prl,superscriptaddress,reprint]{revtex4-1}
\usepackage{graphicx}
\usepackage{bm}
\usepackage{color}
\usepackage[USenglish]{babel}
\newcommand{\mlics}{\mathrm{CsLi}}
\newcommand{\mCsCs}{\mathrm{CsCs}}
\begin{document}
	
	
	\title{Heteronuclear Efimov scenario with positive intraspecies scattering lengths}
	
	\author{Juris Ulmanis}
	\author{Stephan H\"afner}
	\author{Rico Pires}
	\author{Eva D. Kuhnle}
	\affiliation{Physikalisches Institut, Universit\"at Heidelberg, Im Neuenheimer Feld 226, 69120 Heidelberg, Germany}
	
	\author{Yujun Wang}
	\affiliation{Department of Physics, Kansas State University, 116 Cardwell Hall, Manhattan, KS 66506, USA}
	
	\author{Chris H. Greene}
	\email[]{chgreene@purdue.edu}
	\affiliation{Department of Physics, Purdue University, West Lafayette, Indiana, 47907-2036, USA }
	
	\author{Matthias Weidem\"uller}
	\email{weidemueller@uni-heidelberg.de}
	\affiliation{Physikalisches Institut, Universit\"at Heidelberg, Im Neuenheimer Feld 226, 69120 Heidelberg, Germany}
	\affiliation{Hefei National Laboratory for Physical Sciences at the Microscale and Department of Modern Physics,
		and CAS Center for Excellence and Synergetic Innovation Center in Quantum Information and Quantum Physics, University of Science and Technology of China, Hefei, Anhui 230026, China} 
	
	\date{\today}
	
	\begin{abstract}
		We investigate theoretically and experimentally the heteronuclear Efimov scenario for a three-body system that consists of two bosons and one distinguishable particle with positive intraspecies scattering lengths. The three-body parameter at the three-body scattering threshold and the scaling factor between consecutive Efimov resonances are found to be controlled by the scattering length between the two bosons, approximately independent of short-range physics. We observe two excited-state Efimov resonances in the three-body recombination spectra of an ultracold mixture of fermionic $^6 $Li and bosonic $^{133} $Cs atoms close to a Li-Cs Feshbach resonance, where the Cs-Cs interaction is positive. Deviation of the obtained scaling factor of 4.0(3) from the universal prediction of 4.9 and the absence of the ground state Efimov resonance shed new light on the interpretation of the universality and the discrete scaling behavior of heteronuclear Efimov physics.

	\end{abstract}
	
	\pacs{Valid PACS appear here}
	
	\maketitle
	
	The quantum mechanical three-body problem is of fundamental importance for universal phenomena in few-body and nuclear physics~\cite{Jensen2004,Braaten2006,Ferlaino2011,Wang2013,Frederico2012}. A particular example is the Efimov scenario~\cite{Efimov1970,Efimov1971,Efimov1973}, which manifests in a three-body system with pairwise resonant interactions as an infinite geometrical progression of bound states, the Efimov states, that follows a discrete scaling law. Such series have been observed in experiments with ultracold homonuclear Bose~\cite{Huang2014}, three-component Fermi~\cite{Williams2009}, and heteronuclear Bose-Fermi~\cite{Pires2014,Tung2014,Ulmanis2015,Ulmanis2016} gases. 
	Furthermore, the excited helium molecule $^4 $He$ _3$, as observed through Coulomb explosion imaging, has been found to accurately obey the predictions for a universal Efimov trimer~\cite{Kunitski2015}.

	The range of validity of the universal scaling behavior in an actual three-body system is governed by the interplay between long- and short-range interactions. Typically, the effect of this interplay is incorporated in a single quantity, the three-body parameter (3BP), which determines the ground state energy of the Efimov spectrum~\cite{Braaten2006,Ferlaino2011}. One of the alternative definitions of the 3BP which is adopted here uses a scattering observable,  namely the value of the scattering length $ a $, at which the ground Efimov state crosses the three-body scattering threshold. 
	The physical origin of the 3BP in a real system can be traced back to short-range forces and depends sensitively on the specific nature of the interactions~\cite{Wang2014a,Wang2015b,Naidon2014a}. As has been shown recently, in homonuclear atomic systems the 3BP is determined by the characteristic length of the van der Waals (vdW) interaction, independent of the molecular details at short range (vdW universality)~\cite{Berninger2011,Wang2012,Roy2013,Wild2012,Gross2009,Gross2010,Huang2014a,Huang2015a}. Similar universality of the 3BP has been proposed for heteronuclear trimers of kind $ BBX $ that consist of two heavy atoms $ B $ resonantly interacting with a lighter one $ X $, with an additional dependence on the intraspecies interaction between the atoms $ B $~\cite{Wang2012d}. Recent experiments with ultracold Li-Rb~\cite{Maier2015} gases have found resonances that agree well with these predictions, while previously observed features in K-Rb~\cite{Barontini2009} have been challenged by new experiments that align with the vdW universality~\cite{Bloom2013,Wacker2016}. 
	
	In this Letter we theoretically and experimentally investigate universal three-body physics for a heteronuclear system $ BBX $, in which the intraspecies scattering length $ a_{BB} $ is positive, and compare it to the case of negative $ a_{BB} $. This comparison is rendered possible by the occurrence of two favorable Feshbach resonances (FR) in the Li-Cs system~\cite{Repp2013,Tung2013,Pires2014a,Ulmanis2015}. For positive $ a_{BB} $ the heavy particles can form a weakly bound dimer $ BB $, with binding energies comparable to those of the Efimov trimer $ BBX $. This molecular channel significantly modifies the standard Efimov scenario and, in some situations, renders the notion of 3BP superfluous even for contact interactions. The role of the 3BP can be captured by a repulsive barrier that is controlled by the intraspecies scattering lengths. 
	Our measurements of CsCsLi Efimov resonances by three-body recombination rates with positive Cs-Cs scattering length $ a_\mCsCs $ demonstrate the absence of the ground state Efimov resonance, which is predicted to asymptotically converge to the CsCs$ +$Li threshold. In contrast, excited Efimov states are observed as resonances, qualitatively agreeing well with the predictions of the zero-range theory and the spinless vdW model. These results shed new light on the interpretation of the 3BP and the notion of universality in the heteronuclear Efimov scenario.

	\begin{figure}
		\centering
		\includegraphics[width=1\linewidth]{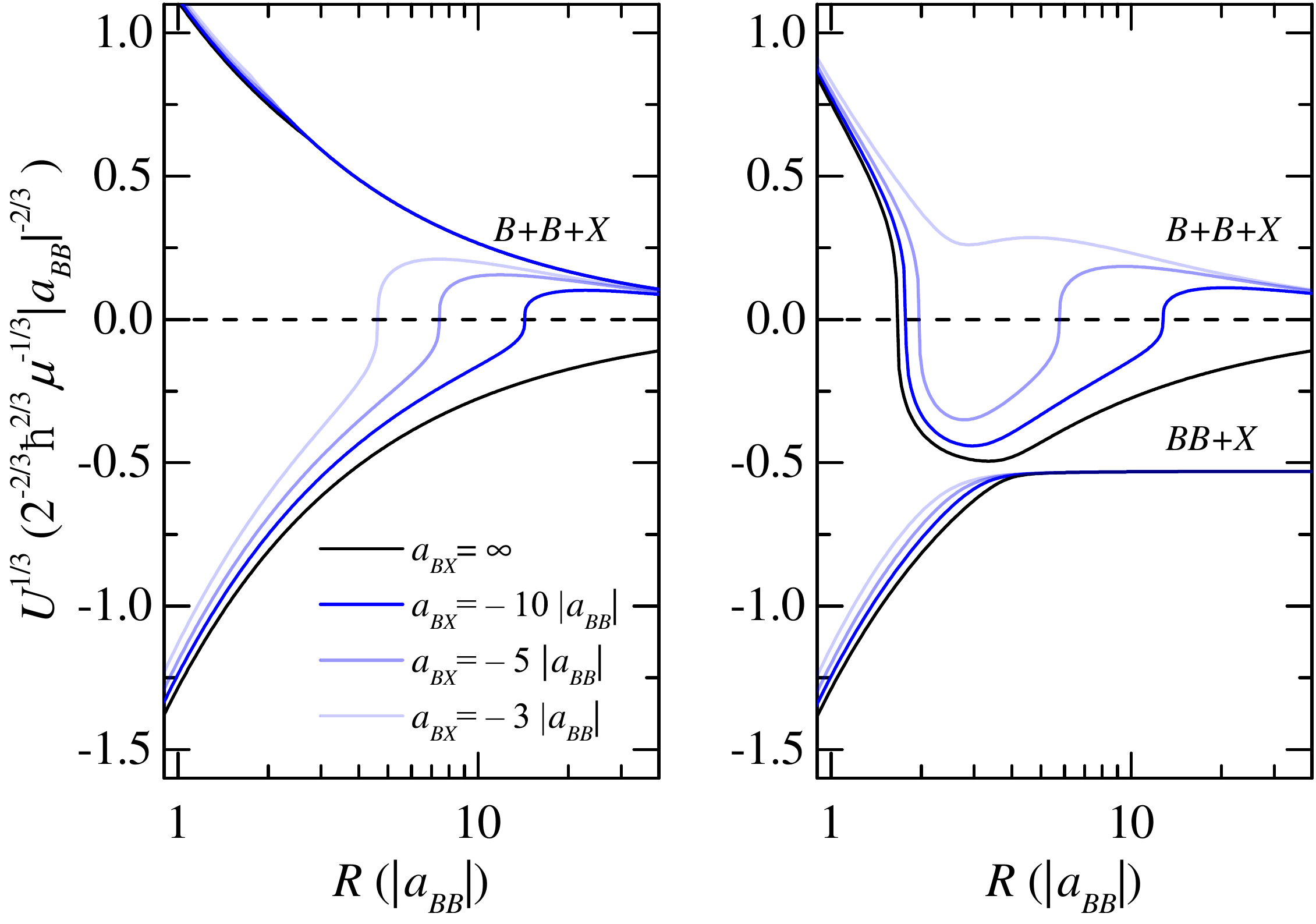}
		\caption[bound state energies]{(color online) Adiabatic hyperspherical potentials $ U $ for a $ BBX $ system with a mass ratio of $ m_B/m_X =22.1$ and $ a_{BB}=-1$ (left panel) and $ a_{BB}=1 $ (right panel) in the universal zero-range model.  The solid curves correspond to interspecies scattering length $ a_{BX} $ with respective values $ \infty $, $ -10 $, $ -5 $, $ -3 $ in units of $ \left|a_{BB}\right| $. The energy unit in this figure is $\hbar^2 / 2 \mu a_{BB}^2$, where $\mu$ is the three-body reduced mass.
		}
		\label{fig:Cs2Li_Pot2}
	\end{figure}

	To qualitatively illustrate the novel features that arise due to the intraspecies interaction, let us first investigate the mass-imbalanced three-body problem assuming pairwise contact interactions (universal zero-range theory). Depending on their signs four different regimes can be defined. Here we consider and compare two so far experimentally relevant cases, which correspond to interspecies scattering length $ a_{BX}<0 $. Adiabatic hyperspherical potential curves with different signs of the intraspecies scattering length $ a_{BB} $ are shown in Fig.~\ref{fig:Cs2Li_Pot2}. If both scattering lengths are negative (Fig.~\ref{fig:Cs2Li_Pot2}, left panel), the classical Efimov scenario is recovered for $ a_{BX}\rightarrow\infty $, with a hyperspherical potential that is proportional to $ -1/R^{2} $, where $ R $ is the hyperradius, and supports an infinite number of bound states. For a finite $ a_{BX} $ a potential barrier forms around $ R\sim 2|a_{BX}| $, leading to quasi-bound states that, by crossing the three-body dissociation threshold, generate three-body recombination resonances in the atom loss-rate spectra. As is the case in any zero-range theory, the energy spectrum has to be regularized with a 3BP, because the attractive inverse square potential is otherwise too singular at $ R\rightarrow 0$.
	
	The situation is significantly different for $ a_{BB}>0 $ (Fig.~\ref{fig:Cs2Li_Pot2}, right panel), where a Feshbach dimer exists with binding energy $ E_{BB}$. The three-body channel $ BB+X $, representing asymptotically the dimer $ BB $ and the particle $ X $, separates the three-body energy spectrum into two branches. The lower Efimov branch, with energy $ E<E_{BB} $ and a hyperspherical potential that asymptotically connects to the $ BB+X $ channel, follows the classical Efimov scenario with a diverging ground state of the energy spectrum prior to regularization. The upper Efimov branch with energy $ E>E_{BB} $ displays qualitatively different behavior. It supports a repulsive potential barrier around $ R\sim 2a_{BB} $ that is roughly independent of the value of $ a_{BX} $. In the resonant limit ($ a_{BX}\rightarrow\infty $) an Efimov potential of $ \propto-1/R^{2} $ character is recovered, but in the adiabatic approximation no regularization is required since the energy spectrum is well-defined~\footnote{In fact a Landau-Zener estimate shows that for $a_{BX}$ near unitarity, these potentials are midway between the diabatic and adiabatic limits for Cs-Cs-Li, but for $|a_{BX}|<10|a_{BB}|$ they are closer to the adiabatic limit.}. For finite $ a_{BX} $, similar to the $a_{BB}<0$ case, an effective potential barrier forms around $ R\sim2|a_{BX}| $ that can lead to recombination resonances. Note that up to this point no other assumptions than the contact two-body interactions and the adiabaticity of hyperspherical potentials have been made.

	\begin{figure}[b]
		\centering
		\includegraphics[width=1\linewidth]{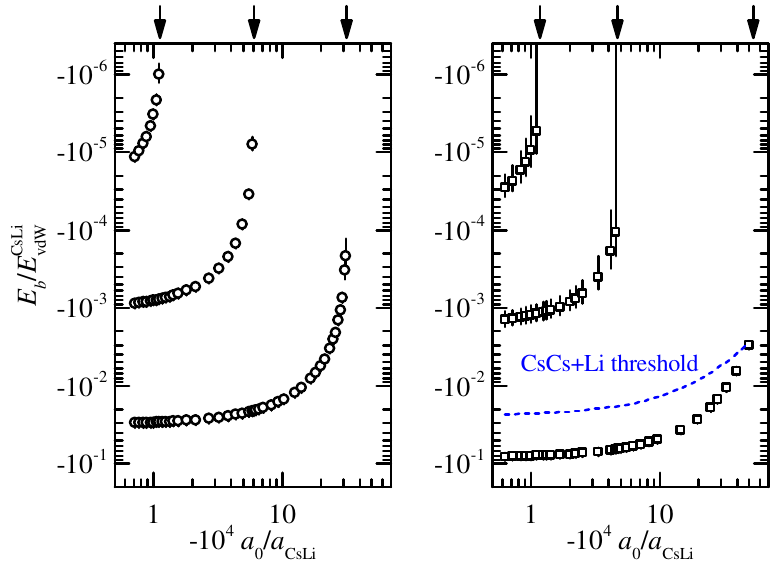}
		\caption[calculated binding energies]{(color online) Calculated CsCsLi energy spectra for the three deepest Efimov states for experimental scattering lengths that correspond to the 843~G (circles, left panel) and 889~G (squares, right panel) Li-Cs Feshbach resonances and $ a_\mCsCs \approx-1500\, a_0$ and $ a_\mCsCs \approx190\, a_0$, respectively~\footnote{Here we use van der Waals energy units, in which $ E^\mlics_\mathrm{vdW}= \hbar^2/(2\mu r_\mathrm{vdW}^2)\approx 156h\ \mathrm{MHz}$, where $ r_\mathrm{vdW}\approx45\, a_0 $ and $ \mu $ denotes the reduced mass.}. The atom-dimer scattering threshold CsCs$ + $Li for the latter case is shown as a dashed blue line. Arrows indicate the positions at which the Efimov states would either cross the Cs$ + $Cs$ + $Li scattering threshold or become predissociative into a CsCs$ + $Li state. The error bars represent the width of the corresponding Efimov state. Missing error bars indicate a width that is smaller than the symbol size.
		}
		\label{fig:E_bind}
	\end{figure}
	
	To study the differences between the scenarios with positive and negative $ a_{BB} $, and the two Efimov branches more quantitatively, we employ the spinless vdW theory~\cite{Wang2012d} and use the CsCsLi system as an example. This approach consists of numerically solving the three-body scattering problem in the hyperspherical formalism for two-body interactions that are modeled with single-channel Lennard-Jones potentials. The scattering lengths $ a_\mCsCs $ and  $ a_\mlics $, as well as their relation $ a_\mCsCs(a_\mlics) $ for the experimentally employed FRs are set by the depth of the respective potentials~\footnote{ The Li-Cs scattering length is inferred from the radio-frequency association of LiCs Feshbach dimers around 843~G and 889~G and a coupled channels calculation~\cite{Ulmanis2015}.  The corresponding Cs-Cs scattering lengths are obtained using the parametrization given in Ref.~\cite{Berninger2013}.}. The adiabatic hyperspherical potentials qualitatively agree with the zero-range theory. Thus, we use them to extract the three-body energy spectra of the three deepest Efimov states that are shown in Fig.~\ref{fig:E_bind}. For negative $ a_\mCsCs $ the classical behavior is recovered, where the states successively disappear through the three-body scattering threshold into the scattering continuum as $ a_\mlics $ is tuned from large to small values. For positive $ a_\mCsCs $ the manifestation of the two Efimov branches, separated by the CsCs$ + $Li threshold, becomes evident. The three-body states from the upper Efimov branch follow the classical scenario, while the state from the lower Efimov branch merges with the atom-dimer threshold and disappears in the atom-dimer scattering continuum. Consequently, it can be expected that the lower-branch state is less likely to produce a resonant feature in the three-body recombination.

	These predictions are tested experimentally for the system of ultracold Li and Cs gases in a setup that is similar to the one presented in Ref.~\cite{Repp2013,Pires2014}. 	In brief, we prepare a mixture of bosonic $ ^{133} $Cs atoms in the $ \left| f,m_f\right\rangle =\left| 3,3\right\rangle  $ magnetic spin state and fermionic $ ^6 $Li atoms in the $\left| 1/2,-1/2\right\rangle  $ state, close to the 889~G Li-Cs FR~\cite{Repp2013,Pires2014a,Tung2013,Ulmanis2015}, where $ a_\mCsCs\approx+190\, a_0 $~\cite{Berninger2013}.  Here $ f $ and $ m_f $ correspond to the total atomic angular momentum and to its projection, respectively. During the measurements the mixture is confined in a bichromatic optical dipole trap that allows us to reach temperatures around 100 nK for each of the species, while retaining sufficient spatial overlap of the gas clouds~\cite{Ulmanis2016}. 
	
	The three-body recombination rates are measured as a function of the external magnetic field, similar to our previous work~\cite{Pires2014,Ulmanis2016}. 
	The loss coefficient $L_3$ for Li-Cs-Cs loss is obtained by numerically fitting the spatially integrated coupled rate equations
	\begin{eqnarray}
		\dot{n}_{\mathrm{Li}} &=& -\alpha_{\mathrm{Li}}{n}_{\mathrm{Li}} -L_3^{} {n}_{\mathrm{Li}} {n}_{\mathrm{Cs}}^{2}
		\label{lossrates1}
		\\
		\dot{n}_{\mathrm{Cs}} &=& -\alpha_{\mathrm{Cs}}{n}_{\mathrm{Cs}} - 2L_3^{} n_{\mathrm{Li}} {n}_{\mathrm{Cs}}^{2} - L_3^{\mathrm{Cs}}{n}_{\mathrm{Cs}}^{3}
		\label{lossrates2}
	\end{eqnarray}
	to the time evolution of the number of atoms in the dipole trap, where $ {n}_{\mathrm{Li}} $ and $ {n}_{\mathrm{Cs}} $ are the densities of Li and Cs atoms, respectively~\footnote{Note that this definition of $ L_3 $ differs by a factor of two from the one that was used in Ref.~\cite{Pires2014}.}. The values of $\alpha_{\mathrm{Li}}$, $\alpha_{\mathrm{Cs}}$  correspond to the background loss rate of each species in the trap and $L_3^{\mathrm{Cs}}$ to the three-body loss coefficient of a pure Cs sample at the same conditions. These parameters are determined in independent measurements, thus the three-body recombination rate constant $L_3^{}$ for Cs$ + $Cs$ + $Li  collisions and the initial atom numbers are the only free fitting parameters. The extraction of $L_3^{}$ is limited by intraspecies loss for the experimentally employed parameters in the range $ -10^4\,a_0/a_\mlics \gtrsim 10$  for the data taken at our lowest temperatures. The error bars are obtained by bootstrapping and represent one standard deviation of the resampled population distributions. Under typical experimental conditions the systematic error of the absolute value of $ L_3 $ is estimated to be a factor of three, caused by uncertainties in the determination of the atom cloud temperatures, densities, overlap and trap frequencies. 
	
	No significant increase of temperature during the entire hold time is observed~\cite{Ulmanis2016}, therefore we omit the recombinational heating~\cite{Weber2003} in our fitting model.  The magnetic field stability is around 16~mG (one standard deviation) resulting from long-term drifts, residual field curvature along the long axis of the cigar-shaped trap and calibration uncertainties.

	\begin{figure}
		\centering
		\includegraphics[width=0.97\linewidth]{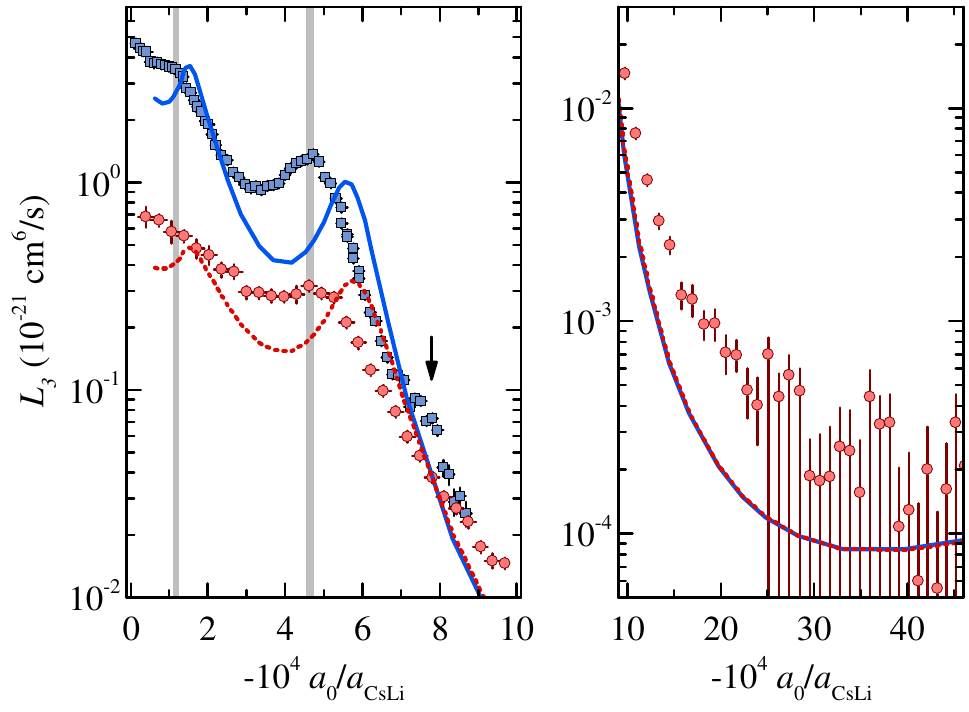}
		\caption[843 G L3]{(color online) Experimental and theoretical Cs-Cs-Li three-body recombination rate spectra at a temperature of 120~nK (blue squares, blue solid line) and 320~nK (red circles, red dotted line) for $ a_\mCsCs\approx190\, a_0 $. The error bars represent the statistical errors from bootstrapping, magnetic field uncertainties, and technical limit due to Cs-Cs-Cs three-body losses. The grayed areas correspond to the zero-temperature Efimov resonance positions, which are extracted from the calculated trimer energy spectra (see text). The arrow in the left panel indicates the approximate position of a possible four-body resonance.}
		\label{fig:L3_comp_889G}
	\end{figure}

	The measured three-body recombination rate spectra around the Li-Cs FR at 889~G ($ a_\mCsCs\approx+190\, a_0 $) are shown in Fig.~\ref{fig:L3_comp_889G} for two different temperature regimes. 
	Two consecutive CsCsLi Efimov resonances are evident. The first recombination feature emerges at a scattering length around $-2000\ a_0$, which is about a factor of six larger than for negative $ a_\mCsCs$~\cite{Tung2014,Pires2014,Ulmanis2015}. This drastic change in the 3BP is consistent with the presented theoretical picture, where the most deeply bound Efimov state disappears into the atom-dimer scattering continuum before reaching the three-body continuum. Thus we assign the first recombination feature to the first excited Efimov resonance and the upper Efimov branch. The observed general power law of the $ L^{}_3 $ scaling with $ a_\mlics $ and $ a_\mCsCs $ qualitatively agrees with scaling laws of the three-body collision rates near overlapping FRs~\cite{DIncao2009a,tbp}.
	
	The ground state recombination feature is missing in the recombination spectra of the vdW theory, which predicts only two resonances that agree with the experimentally determined features. Despite the qualitative agreement, there is a shift between the theoretical and experimental resonance positions. The shift may be an indication of the multichannel nature of the FR~\cite{Wang2014a,Sorensen2012} that is used for tuning $ a_\mlics $ experimentally, but neglected in our spinless theory. More detailed investigations will be required to identify the origin of this quantitative discrepancy. The spinless vdW model shows quantitative agreement with our measurements close to the 843~G Li-Cs FR, featuring negative $ a_\mCsCs $~\cite{tbp}.

	In order to compare our results with the commonly used analysis of Efimov universality in the terms of 3BPs and scaling factors, we extract these quantities from the spinless vdW model. The resonance positions cannot be directly fitted from the experimental loss-rate data, since the excited resonance features are all already located into the unitarity limited regime, and, thus, effects of finite temperature would have to be included (for negative intraspecies scattering length, see Refs.~\cite{Petrov2015,Mikkelsen2015,Ulmanis2016}).
	Therefore, encouraged by the good agreement between the experimental findings and the vdW model, we extract Efimov resonance positions $ a_{-,\mlics}^{(n)} $ from the calculated trimer energy spectra as the average value of the two numerical grid points, between which the three-body states merge with the scattering continuum (see Fig.~\ref{fig:E_bind}). In this way the influence of finite temperature can be safely neglected. This procedure yields $a^{(1)}_{-,\mlics}=-2150(50)\, a_0$ and $a^{(2)}_{-,\mlics}=-8500(500)\, a_0 $, where the value in brackets gives one half the step size of the local grid. The corresponding effective scaling factor at the dissociation threshold is $ \lambda=4.0(3) $.  (For the determination of resonance positions and the effective scaling factor that uses Gaussian profiles with linear background see Supplemental material).
	
	It is instructive to compare this finding to previous theoretical predictions, which provided a scaling factor of 4.9 for consecutive CsCsLi Efimov states~\cite{DIncao2006,Braaten2006,Helfrich2010a}. These theories assumed resonant interspecies interactions  $ (a_\mlics\rightarrow\infty)$ and negligible intraspecies interactions $(a_\mCsCs=0)$. The former assumption is well fulfilled for the observed resonance positions of $a^{(1)}_{-,\mlics}$ and $a^{(2)}_{-,\mlics}$, which are much larger than the vdW lengths of $45\, a_0$ and $ 101\, a_0 $ for Li-Cs and Cs-Cs atom pairs, respectively. Furthermore, since $ a_\mCsCs\approx+190\, a_0$, the resonances can be expected to follow the resonant limit, in which $ a_\mlics $ is the largest length scale. Nevertheless, deviation from the universal prediction is evident, similar to the case of negative $ a_\mlics $~\cite{Ulmanis2016}. Thus, the description of an actual heteronuclear three-body system with finite scattering lengths in terms of a single, constant scaling factor is questionable even in the zero-range theory. The use of an Efimov-period-dependent scaling factor $ \lambda_{n} $ seems to be more appropriate. The calculated $ \lambda_{n} $ between higher lying states approach the universal constant, as expected.
	
	Our measurements in the CsCsLi system show that the 3BP at the $ B+B+X $ threshold depends on the intraspecies interaction. The hyperspherical adiabatic potentials within the zero-range theory suggest that two Efimov branches exist, where the upper one supports an effective potential barrier, which is largely determined by $ a_{BB} $. A potential barrier with similar properties has been theoretically discussed previously for light-light-heavy systems~\cite{Wang2012d}. Thus it is possible that the 3BP on the three-body scattering threshold in a heteronuclear three-atom system that features positive $ a_{BB} $ and negative $ a_{BX} $ is independent of vdW forces. Since such a barrier already appears in the zero-range theory, it is reasonable to assume that not only atomic but any heteronuclear system with short-range interactions will possess a 3BP that is uniquely determined by $ a_{BB} $. The observed absence of the CsCsLi ground state Efimov resonance, where it is supported by the potential of the lower Efimov branch that asymptotically merges with the CsCs$+ $Li threshold, is consistent with this theoretical picture. Observation of an atom-dimer feature around $ a_\mlics\approx-200\, a_0 $ in the CsCs$+ $Li channel would strengthen this interpretation and shed new light on universal few-body scenarios in mixed systems. These experiments would require extension of the current preparation procedures to an ultracold mixture of Li atoms and Cs$ _2 $ Feshbach molecules. The properties of possible atom-dimer vibrational relaxation features is an intriguing question in itself, since no effective potential barrier is evident in the present calculation. The existence of two Efimov branches may also help to explain the breakdown of universal relationships (which assume $ a_{BB}\rightarrow0 $) between resonance and interference features in K-Rb mixtures~\cite{Bloom2013,Helfrich2010}.

	The slight increase of the three-body loss rates around  $a_\mlics\approx0.6\ a^{(1)}_{-,\mlics} $ for the data set at 120~nK, close to the 889~G FR, may indicate a Cs-Cs-Cs-Li four-body recombination resonance (see arrow in Fig.~\ref{fig:L3_comp_889G}). The position is in good agreement with the theoretical prediction of  $0.55\ a^{(1)}_{-} $~\cite{Blume2014}. The second resonance, which is expected to appear at  $0.91\ a^{(1)}_{-} $, would already overlap with the three-body recombination resonance. 
	However, at the current stage of the experiment, we cannot unambiguously validate contributions from four-body processes, since an analogous fitting model to the one in Eq.~(\ref{lossrates1}) and~(\ref{lossrates2}) with additional four-body loss terms does not improve the overall quality of the fit and does not result in pronounced features in the four-body loss rate spectrum.
	
	In conclusion, we have experimentally and theoretically investigated the behavior of the 3BP in a heteronuclear three-body system with positive and negative intraspecies interactions. For $ a_{BB}<0 $ the classical Efimov scenario is recovered, where the 3BP is determined by short-range forces. In contrast, for $ a_{BB}>0 $ two Efimov branches exist, with different microscopic mechanisms governing the value of the 3BP. The observation of two consecutive excited state Efimov resonances and the absence of the ground state resonance in the Cs-Cs-Li system for positive $ a_\mCsCs $ is consistent not only with the Efimov branches, but also suggests a scaling factor and a 3BP that depend primarily on the values of the intra- and interspecies scattering lengths.

	\begin{acknowledgments}
		We are grateful to P. Giannakeas, R. Grimm, S. Jochim, A.L. de Oliveira, S. Whitlock, and N. Zinner for fruitful discussions. This work is supported in part by the Heidelberg Center for Quantum Dynamics. J.U., S.H. and R.P. acknowledge support by the IMPRS-QD, J.U. by the DAAD. E. K. is indebted to the Baden- W\"urttemberg Stiftung for the financial support of this research project by the Eliteprogramme for Postdocs. Y.W. acknowledges support by an AFOSR-MURI and Department of Physics, Kansas State University, and C.H.G. from the Binational Science Foundation, Grant No. 2012504. This work was supported in part by the Deutsche Forschungsgemeinschaft under Project No. WE2661/11-1 and Collaborative Research Centre “SFB 1225 (ISOQUANT).”
	\end{acknowledgments}
	
	\IfFileExists{T:/quantdyn/Literature/Mixtures/Mixtures.bib}{\bibliography{T:/quantdyn/Literature/Mixtures/Mixtures}}{} 

\begin{thebibliography}{48}%
\makeatletter
\providecommand \@ifxundefined [1]{%
 \@ifx{#1\undefined}
}%
\providecommand \@ifnum [1]{%
 \ifnum #1\expandafter \@firstoftwo
 \else \expandafter \@secondoftwo
 \fi
}%
\providecommand \@ifx [1]{%
 \ifx #1\expandafter \@firstoftwo
 \else \expandafter \@secondoftwo
 \fi
}%
\providecommand \natexlab [1]{#1}%
\providecommand \enquote  [1]{``#1''}%
\providecommand \bibnamefont  [1]{#1}%
\providecommand \bibfnamefont [1]{#1}%
\providecommand \citenamefont [1]{#1}%
\providecommand \href@noop [0]{\@secondoftwo}%
\providecommand \href [0]{\begingroup \@sanitize@url \@href}%
\providecommand \@href[1]{\@@startlink{#1}\@@href}%
\providecommand \@@href[1]{\endgroup#1\@@endlink}%
\providecommand \@sanitize@url [0]{\catcode `\\12\catcode `\$12\catcode
  `\&12\catcode `\#12\catcode `\^12\catcode `\_12\catcode `\%12\relax}%
\providecommand \@@startlink[1]{}%
\providecommand \@@endlink[0]{}%
\providecommand \url  [0]{\begingroup\@sanitize@url \@url }%
\providecommand \@url [1]{\endgroup\@href {#1}{\urlprefix }}%
\providecommand \urlprefix  [0]{URL }%
\providecommand \Eprint [0]{\href }%
\providecommand \doibase [0]{http://dx.doi.org/}%
\providecommand \selectlanguage [0]{\@gobble}%
\providecommand \bibinfo  [0]{\@secondoftwo}%
\providecommand \bibfield  [0]{\@secondoftwo}%
\providecommand \translation [1]{[#1]}%
\providecommand \BibitemOpen [0]{}%
\providecommand \bibitemStop [0]{}%
\providecommand \bibitemNoStop [0]{.\EOS\space}%
\providecommand \EOS [0]{\spacefactor3000\relax}%
\providecommand \BibitemShut  [1]{\csname bibitem#1\endcsname}%
\let\auto@bib@innerbib\@empty
\bibitem [{\citenamefont {Jensen}\ \emph {et~al.}(2004)\citenamefont {Jensen},
  \citenamefont {Riisager}, \citenamefont {Fedorov},\ and\ \citenamefont
  {Garrido}}]{Jensen2004}%
  \BibitemOpen
  \bibfield  {author} {\bibinfo {author} {\bibfnamefont {A.~S.}\ \bibnamefont
  {Jensen}}, \bibinfo {author} {\bibfnamefont {K.}~\bibnamefont {Riisager}},
  \bibinfo {author} {\bibfnamefont {D.~V.}\ \bibnamefont {Fedorov}}, \ and\
  \bibinfo {author} {\bibfnamefont {E.}~\bibnamefont {Garrido}},\ }\href
  {\doibase 10.1103/RevModPhys.76.215} {\bibfield  {journal} {\bibinfo
  {journal} {Rev. Mod. Phys.}\ }\textbf {\bibinfo {volume} {76}},\ \bibinfo
  {pages} {215} (\bibinfo {year} {2004})}\BibitemShut {NoStop}%
\bibitem [{\citenamefont {Braaten}\ and\ \citenamefont
  {Hammer}(2006)}]{Braaten2006}%
  \BibitemOpen
  \bibfield  {author} {\bibinfo {author} {\bibfnamefont {E.}~\bibnamefont
  {Braaten}}\ and\ \bibinfo {author} {\bibfnamefont {H.-W.}\ \bibnamefont
  {Hammer}},\ }\href {\doibase http://dx.doi.org/10.1016/j.physrep.2006.03.001}
  {\bibfield  {journal} {\bibinfo  {journal} {Phys. Rep.}\ }\textbf {\bibinfo
  {volume} {428}},\ \bibinfo {pages} {259} (\bibinfo {year}
  {2006})}\BibitemShut {NoStop}%
\bibitem [{\citenamefont {Ferlaino}\ \emph {et~al.}(2011)\citenamefont
  {Ferlaino}, \citenamefont {Zenesini}, \citenamefont {Berninger},
  \citenamefont {Huang}, \citenamefont {N\"agerl},\ and\ \citenamefont
  {Grimm}}]{Ferlaino2011}%
  \BibitemOpen
  \bibfield  {author} {\bibinfo {author} {\bibfnamefont {F.}~\bibnamefont
  {Ferlaino}}, \bibinfo {author} {\bibfnamefont {A.}~\bibnamefont {Zenesini}},
  \bibinfo {author} {\bibfnamefont {M.}~\bibnamefont {Berninger}}, \bibinfo
  {author} {\bibfnamefont {B.}~\bibnamefont {Huang}}, \bibinfo {author}
  {\bibfnamefont {H.-C.}\ \bibnamefont {N\"agerl}}, \ and\ \bibinfo {author}
  {\bibfnamefont {R.}~\bibnamefont {Grimm}},\ }\href {\doibase
  10.1007/s00601-011-0260-7} {\bibfield  {journal} {\bibinfo  {journal}
  {Few-Body Syst.}\ }\textbf {\bibinfo {volume} {51}},\ \bibinfo {pages} {113}
  (\bibinfo {year} {2011})}\BibitemShut {NoStop}%
\bibitem [{\citenamefont {Wang}\ \emph {et~al.}(2013)\citenamefont {Wang},
  \citenamefont {D'Incao},\ and\ \citenamefont {Esry}}]{Wang2013}%
  \BibitemOpen
  \bibfield  {author} {\bibinfo {author} {\bibfnamefont {Y.}~\bibnamefont
  {Wang}}, \bibinfo {author} {\bibfnamefont {J.~P.}\ \bibnamefont {D'Incao}}, \
  and\ \bibinfo {author} {\bibfnamefont {B.~D.}\ \bibnamefont {Esry}},\ }\href
  {\doibase http://dx.doi.org/10.1016/B978-0-12-408090-4.00001-3} {\emph
  {\bibinfo {title} {Ultracold Few-Body Systems}}},\ edited by\ \bibinfo
  {editor} {\bibfnamefont {P.~R.~B.}\ \bibnamefont {Ennio~Arimondo}}\ and\
  \bibinfo {editor} {\bibfnamefont {C.~C.}\ \bibnamefont {Lin}},\ \bibinfo
  {series} {Advances in Atomic, Molecular, and Optical Physics}, Vol.~\bibinfo
  {volume} {62}\ (\bibinfo  {publisher} {Academic Press},\ \bibinfo {year}
  {2013})\BibitemShut {NoStop}%
\bibitem [{\citenamefont {Frederico}\ \emph {et~al.}(2012)\citenamefont
  {Frederico}, \citenamefont {Delfino}, \citenamefont {Tomio},\ and\
  \citenamefont {Yamashita}}]{Frederico2012}%
  \BibitemOpen
  \bibfield  {author} {\bibinfo {author} {\bibfnamefont {T.}~\bibnamefont
  {Frederico}}, \bibinfo {author} {\bibfnamefont {A.}~\bibnamefont {Delfino}},
  \bibinfo {author} {\bibfnamefont {L.}~\bibnamefont {Tomio}}, \ and\ \bibinfo
  {author} {\bibfnamefont {M.}~\bibnamefont {Yamashita}},\ }\href {\doibase
  http://dx.doi.org/10.1016/j.ppnp.2012.06.001} {\bibfield  {journal} {\bibinfo
   {journal} {Prog. Part. Nucl. Phys.}\ }\textbf {\bibinfo {volume} {67}},\
  \bibinfo {pages} {939} (\bibinfo {year} {2012})}\BibitemShut {NoStop}%
\bibitem [{\citenamefont {Efimov}(1970)}]{Efimov1970}%
  \BibitemOpen
  \bibfield  {author} {\bibinfo {author} {\bibfnamefont {V.}~\bibnamefont
  {Efimov}},\ }\href {\doibase http://dx.doi.org/10.1016/0370-2693(70)90349-7}
  {\bibfield  {journal} {\bibinfo  {journal} {Phys. Lett. B}\ }\textbf
  {\bibinfo {volume} {33}},\ \bibinfo {pages} {563} (\bibinfo {year}
  {1970})}\BibitemShut {NoStop}%
\bibitem [{\citenamefont {Efimov}(1971)}]{Efimov1971}%
  \BibitemOpen
  \bibfield  {author} {\bibinfo {author} {\bibfnamefont {V.}~\bibnamefont
  {Efimov}},\ }\href@noop {} {\bibfield  {journal} {\bibinfo  {journal} {Sov.
  J. Nuc. Phys.}\ }\textbf {\bibinfo {volume} {12}},\ \bibinfo {pages} {589}
  (\bibinfo {year} {1971})}\BibitemShut {NoStop}%
\bibitem [{\citenamefont {Efimov}(1973)}]{Efimov1973}%
  \BibitemOpen
  \bibfield  {author} {\bibinfo {author} {\bibfnamefont {V.}~\bibnamefont
  {Efimov}},\ }\href {\doibase http://dx.doi.org/10.1016/0375-9474(73)90510-1}
  {\bibfield  {journal} {\bibinfo  {journal} {Nucl. Phys. A}\ }\textbf
  {\bibinfo {volume} {210}},\ \bibinfo {pages} {157} (\bibinfo {year}
  {1973})}\BibitemShut {NoStop}%
\bibitem [{\citenamefont {Huang}\ \emph
  {et~al.}(2014{\natexlab{a}})\citenamefont {Huang}, \citenamefont
  {Sidorenkov}, \citenamefont {Grimm},\ and\ \citenamefont
  {Hutson}}]{Huang2014}%
  \BibitemOpen
  \bibfield  {author} {\bibinfo {author} {\bibfnamefont {B.}~\bibnamefont
  {Huang}}, \bibinfo {author} {\bibfnamefont {L.~A.}\ \bibnamefont
  {Sidorenkov}}, \bibinfo {author} {\bibfnamefont {R.}~\bibnamefont {Grimm}}, \
  and\ \bibinfo {author} {\bibfnamefont {J.~M.}\ \bibnamefont {Hutson}},\
  }\href {\doibase 10.1103/PhysRevLett.112.190401} {\bibfield  {journal}
  {\bibinfo  {journal} {Phys. Rev. Lett.}\ }\textbf {\bibinfo {volume} {112}},\
  \bibinfo {pages} {190401} (\bibinfo {year} {2014}{\natexlab{a}})}\BibitemShut
  {NoStop}%
\bibitem [{\citenamefont {Williams}\ \emph {et~al.}(2009)\citenamefont
  {Williams}, \citenamefont {Hazlett}, \citenamefont {Huckans}, \citenamefont
  {Stites}, \citenamefont {Zhang},\ and\ \citenamefont
  {O'Hara}}]{Williams2009}%
  \BibitemOpen
  \bibfield  {author} {\bibinfo {author} {\bibfnamefont {J.~R.}\ \bibnamefont
  {Williams}}, \bibinfo {author} {\bibfnamefont {E.~L.}\ \bibnamefont
  {Hazlett}}, \bibinfo {author} {\bibfnamefont {J.~H.}\ \bibnamefont
  {Huckans}}, \bibinfo {author} {\bibfnamefont {R.~W.}\ \bibnamefont {Stites}},
  \bibinfo {author} {\bibfnamefont {Y.}~\bibnamefont {Zhang}}, \ and\ \bibinfo
  {author} {\bibfnamefont {K.~M.}\ \bibnamefont {O'Hara}},\ }\href {\doibase
  10.1103/PhysRevLett.103.130404} {\bibfield  {journal} {\bibinfo  {journal}
  {Phys. Rev. Lett.}\ }\textbf {\bibinfo {volume} {103}},\ \bibinfo {pages}
  {130404} (\bibinfo {year} {2009})}\BibitemShut {NoStop}%
\bibitem [{\citenamefont {Pires}\ \emph
  {et~al.}(2014{\natexlab{a}})\citenamefont {Pires}, \citenamefont {Ulmanis},
  \citenamefont {H\"afner}, \citenamefont {Repp}, \citenamefont {Arias},
  \citenamefont {Kuhnle},\ and\ \citenamefont {Weidem\"uller}}]{Pires2014}%
  \BibitemOpen
  \bibfield  {author} {\bibinfo {author} {\bibfnamefont {R.}~\bibnamefont
  {Pires}}, \bibinfo {author} {\bibfnamefont {J.}~\bibnamefont {Ulmanis}},
  \bibinfo {author} {\bibfnamefont {S.}~\bibnamefont {H\"afner}}, \bibinfo
  {author} {\bibfnamefont {M.}~\bibnamefont {Repp}}, \bibinfo {author}
  {\bibfnamefont {A.}~\bibnamefont {Arias}}, \bibinfo {author} {\bibfnamefont
  {E.~D.}\ \bibnamefont {Kuhnle}}, \ and\ \bibinfo {author} {\bibfnamefont
  {M.}~\bibnamefont {Weidem\"uller}},\ }\href {\doibase
  10.1103/PhysRevLett.112.250404} {\bibfield  {journal} {\bibinfo  {journal}
  {Phys. Rev. Lett.}\ }\textbf {\bibinfo {volume} {112}},\ \bibinfo {pages}
  {250404} (\bibinfo {year} {2014}{\natexlab{a}})}\BibitemShut {NoStop}%
\bibitem [{\citenamefont {Tung}\ \emph {et~al.}(2014)\citenamefont {Tung},
  \citenamefont {Jim\'enez-Garc\'ia}, \citenamefont {Johansen}, \citenamefont
  {Parker},\ and\ \citenamefont {Chin}}]{Tung2014}%
  \BibitemOpen
  \bibfield  {author} {\bibinfo {author} {\bibfnamefont {S.-K.}\ \bibnamefont
  {Tung}}, \bibinfo {author} {\bibfnamefont {K.}~\bibnamefont
  {Jim\'enez-Garc\'ia}}, \bibinfo {author} {\bibfnamefont {J.}~\bibnamefont
  {Johansen}}, \bibinfo {author} {\bibfnamefont {C.~V.}\ \bibnamefont
  {Parker}}, \ and\ \bibinfo {author} {\bibfnamefont {C.}~\bibnamefont
  {Chin}},\ }\href {\doibase 10.1103/PhysRevLett.113.240402} {\bibfield
  {journal} {\bibinfo  {journal} {Phys. Rev. Lett.}\ }\textbf {\bibinfo
  {volume} {113}},\ \bibinfo {pages} {240402} (\bibinfo {year}
  {2014})}\BibitemShut {NoStop}%
\bibitem [{\citenamefont {Ulmanis}\ \emph {et~al.}(2015)\citenamefont
  {Ulmanis}, \citenamefont {H\"afner}, \citenamefont {Pires}, \citenamefont
  {Kuhnle}, \citenamefont {Weidem\"uller},\ and\ \citenamefont
  {Tiemann}}]{Ulmanis2015}%
  \BibitemOpen
  \bibfield  {author} {\bibinfo {author} {\bibfnamefont {J.}~\bibnamefont
  {Ulmanis}}, \bibinfo {author} {\bibfnamefont {S.}~\bibnamefont {H\"afner}},
  \bibinfo {author} {\bibfnamefont {R.}~\bibnamefont {Pires}}, \bibinfo
  {author} {\bibfnamefont {E.~D.}\ \bibnamefont {Kuhnle}}, \bibinfo {author}
  {\bibfnamefont {M.}~\bibnamefont {Weidem\"uller}}, \ and\ \bibinfo {author}
  {\bibfnamefont {E.}~\bibnamefont {Tiemann}},\ }\href
  {http://stacks.iop.org/1367-2630/17/i=5/a=055009} {\bibfield  {journal}
  {\bibinfo  {journal} {New J. Phys.}\ }\textbf {\bibinfo {volume} {17}},\
  \bibinfo {pages} {055009} (\bibinfo {year} {2015})}\BibitemShut {NoStop}%
\bibitem [{\citenamefont {Ulmanis}\ \emph {et~al.}(2016)\citenamefont
  {Ulmanis}, \citenamefont {H\"afner}, \citenamefont {Pires}, \citenamefont
  {Werner}, \citenamefont {Petrov}, \citenamefont {Kuhnle},\ and\ \citenamefont
  {Weidem\"uller}}]{Ulmanis2016}%
  \BibitemOpen
  \bibfield  {author} {\bibinfo {author} {\bibfnamefont {J.}~\bibnamefont
  {Ulmanis}}, \bibinfo {author} {\bibfnamefont {S.}~\bibnamefont {H\"afner}},
  \bibinfo {author} {\bibfnamefont {R.}~\bibnamefont {Pires}}, \bibinfo
  {author} {\bibfnamefont {F.}~\bibnamefont {Werner}}, \bibinfo {author}
  {\bibfnamefont {D.~S.}\ \bibnamefont {Petrov}}, \bibinfo {author}
  {\bibfnamefont {E.~D.}\ \bibnamefont {Kuhnle}}, \ and\ \bibinfo {author}
  {\bibfnamefont {M.}~\bibnamefont {Weidem\"uller}},\ }\href {\doibase
  10.1103/PhysRevA.93.022707} {\bibfield  {journal} {\bibinfo  {journal} {Phys.
  Rev. A}\ }\textbf {\bibinfo {volume} {93}},\ \bibinfo {pages} {022707}
  (\bibinfo {year} {2016})}\BibitemShut {NoStop}%
\bibitem [{\citenamefont {Kunitski}\ \emph {et~al.}(2015)\citenamefont
  {Kunitski}, \citenamefont {Zeller}, \citenamefont {Voigtsberger},
  \citenamefont {Kalinin}, \citenamefont {Schmidt}, \citenamefont
  {Sch\"{o}ffler}, \citenamefont {Czasch}, \citenamefont {Sch\"{o}llkopf},
  \citenamefont {Grisenti}, \citenamefont {Jahnke}, \citenamefont {Blume},\
  and\ \citenamefont {D\"{o}rner}}]{Kunitski2015}%
  \BibitemOpen
  \bibfield  {author} {\bibinfo {author} {\bibfnamefont {M.}~\bibnamefont
  {Kunitski}}, \bibinfo {author} {\bibfnamefont {S.}~\bibnamefont {Zeller}},
  \bibinfo {author} {\bibfnamefont {J.}~\bibnamefont {Voigtsberger}}, \bibinfo
  {author} {\bibfnamefont {A.}~\bibnamefont {Kalinin}}, \bibinfo {author}
  {\bibfnamefont {L.~P.~H.}\ \bibnamefont {Schmidt}}, \bibinfo {author}
  {\bibfnamefont {M.}~\bibnamefont {Sch\"{o}ffler}}, \bibinfo {author}
  {\bibfnamefont {A.}~\bibnamefont {Czasch}}, \bibinfo {author} {\bibfnamefont
  {W.}~\bibnamefont {Sch\"{o}llkopf}}, \bibinfo {author} {\bibfnamefont
  {R.~E.}\ \bibnamefont {Grisenti}}, \bibinfo {author} {\bibfnamefont
  {T.}~\bibnamefont {Jahnke}}, \bibinfo {author} {\bibfnamefont
  {D.}~\bibnamefont {Blume}}, \ and\ \bibinfo {author} {\bibfnamefont
  {R.}~\bibnamefont {D\"{o}rner}},\ }\href {\doibase 10.1126/science.aaa5601}
  {\bibfield  {journal} {\bibinfo  {journal} {Science}\ }\textbf {\bibinfo
  {volume} {348}},\ \bibinfo {pages} {551} (\bibinfo {year}
  {2015})}\BibitemShut {NoStop}%
\bibitem [{\citenamefont {Wang}\ and\ \citenamefont
  {Julienne}(2014)}]{Wang2014a}%
  \BibitemOpen
  \bibfield  {author} {\bibinfo {author} {\bibfnamefont {Y.}~\bibnamefont
  {Wang}}\ and\ \bibinfo {author} {\bibfnamefont {P.~S.}\ \bibnamefont
  {Julienne}},\ }\href {\doibase doi:10.1038/nphys3071} {\bibfield  {journal}
  {\bibinfo  {journal} {Nat. Phys.}\ }\textbf {\bibinfo {volume} {10}},\
  \bibinfo {pages} {768} (\bibinfo {year} {2014})}\BibitemShut {NoStop}%
\bibitem [{\citenamefont {Wang}\ \emph {et~al.}(2015)\citenamefont {Wang},
  \citenamefont {Julienne},\ and\ \citenamefont {Greene}}]{Wang2015b}%
  \BibitemOpen
  \bibfield  {author} {\bibinfo {author} {\bibfnamefont {Y.}~\bibnamefont
  {Wang}}, \bibinfo {author} {\bibfnamefont {P.}~\bibnamefont {Julienne}}, \
  and\ \bibinfo {author} {\bibfnamefont {C.~H.}\ \bibnamefont {Greene}},\ }in\
  \href {http://www.worldscientific.com/doi/abs/10.1142/9789814667746_0002}
  {\emph {\bibinfo {booktitle} {Annual Review of Cold Atoms and Molecules}}},\
  Vol.~\bibinfo {volume} {3},\ \bibinfo {editor} {edited by\ \bibinfo {editor}
  {\bibfnamefont {K.~W.}\ \bibnamefont {Madison}}, \bibinfo {editor}
  {\bibfnamefont {K.}~\bibnamefont {Bongs}}, \bibinfo {editor} {\bibfnamefont
  {L.~D.}\ \bibnamefont {Carr}}, \bibinfo {editor} {\bibfnamefont {A.~M.}\
  \bibnamefont {Rey}}, \ and\ \bibinfo {editor} {\bibfnamefont
  {H.}~\bibnamefont {Zhai}}}\ (\bibinfo  {publisher} {World Scientific
  Publishing},\ \bibinfo {year} {2015})\ Chap.~\bibinfo {chapter} {2}, pp.\
  \bibinfo {pages} {77--134},\ \bibinfo {note} {also at:
  http://arxiv.org/abs/1412.8094v1}\BibitemShut {NoStop}%
\bibitem [{\citenamefont {Naidon}\ \emph {et~al.}(2014)\citenamefont {Naidon},
  \citenamefont {Endo},\ and\ \citenamefont {Ueda}}]{Naidon2014a}%
  \BibitemOpen
  \bibfield  {author} {\bibinfo {author} {\bibfnamefont {P.}~\bibnamefont
  {Naidon}}, \bibinfo {author} {\bibfnamefont {S.}~\bibnamefont {Endo}}, \ and\
  \bibinfo {author} {\bibfnamefont {M.}~\bibnamefont {Ueda}},\ }\href {\doibase
  10.1103/PhysRevLett.112.105301} {\bibfield  {journal} {\bibinfo  {journal}
  {Phys. Rev. Lett.}\ }\textbf {\bibinfo {volume} {112}},\ \bibinfo {pages}
  {105301} (\bibinfo {year} {2014})}\BibitemShut {NoStop}%
\bibitem [{\citenamefont {Berninger}\ \emph {et~al.}(2011)\citenamefont
  {Berninger}, \citenamefont {Zenesini}, \citenamefont {Huang}, \citenamefont
  {Harm}, \citenamefont {N\"agerl}, \citenamefont {Ferlaino}, \citenamefont
  {Grimm}, \citenamefont {Julienne},\ and\ \citenamefont
  {Hutson}}]{Berninger2011}%
  \BibitemOpen
  \bibfield  {author} {\bibinfo {author} {\bibfnamefont {M.}~\bibnamefont
  {Berninger}}, \bibinfo {author} {\bibfnamefont {A.}~\bibnamefont {Zenesini}},
  \bibinfo {author} {\bibfnamefont {B.}~\bibnamefont {Huang}}, \bibinfo
  {author} {\bibfnamefont {W.}~\bibnamefont {Harm}}, \bibinfo {author}
  {\bibfnamefont {H.-C.}\ \bibnamefont {N\"agerl}}, \bibinfo {author}
  {\bibfnamefont {F.}~\bibnamefont {Ferlaino}}, \bibinfo {author}
  {\bibfnamefont {R.}~\bibnamefont {Grimm}}, \bibinfo {author} {\bibfnamefont
  {P.~S.}\ \bibnamefont {Julienne}}, \ and\ \bibinfo {author} {\bibfnamefont
  {J.~M.}\ \bibnamefont {Hutson}},\ }\href {\doibase
  10.1103/PhysRevLett.107.120401} {\bibfield  {journal} {\bibinfo  {journal}
  {Phys. Rev. Lett.}\ }\textbf {\bibinfo {volume} {107}},\ \bibinfo {pages}
  {120401} (\bibinfo {year} {2011})}\BibitemShut {NoStop}%
\bibitem [{\citenamefont {Wang}\ \emph
  {et~al.}(2012{\natexlab{a}})\citenamefont {Wang}, \citenamefont {D'Incao},
  \citenamefont {Esry},\ and\ \citenamefont {Greene}}]{Wang2012}%
  \BibitemOpen
  \bibfield  {author} {\bibinfo {author} {\bibfnamefont {J.}~\bibnamefont
  {Wang}}, \bibinfo {author} {\bibfnamefont {J.~P.}\ \bibnamefont {D'Incao}},
  \bibinfo {author} {\bibfnamefont {B.~D.}\ \bibnamefont {Esry}}, \ and\
  \bibinfo {author} {\bibfnamefont {C.~H.}\ \bibnamefont {Greene}},\ }\href
  {\doibase 10.1103/PhysRevLett.108.263001} {\bibfield  {journal} {\bibinfo
  {journal} {Phys. Rev. Lett.}\ }\textbf {\bibinfo {volume} {108}},\ \bibinfo
  {pages} {263001} (\bibinfo {year} {2012}{\natexlab{a}})}\BibitemShut
  {NoStop}%
\bibitem [{\citenamefont {Roy}\ \emph {et~al.}(2013)\citenamefont {Roy},
  \citenamefont {Landini}, \citenamefont {Trenkwalder}, \citenamefont
  {Semeghini}, \citenamefont {Spagnolli}, \citenamefont {Simoni}, \citenamefont
  {Fattori}, \citenamefont {Inguscio},\ and\ \citenamefont
  {Modugno}}]{Roy2013}%
  \BibitemOpen
  \bibfield  {author} {\bibinfo {author} {\bibfnamefont {S.}~\bibnamefont
  {Roy}}, \bibinfo {author} {\bibfnamefont {M.}~\bibnamefont {Landini}},
  \bibinfo {author} {\bibfnamefont {A.}~\bibnamefont {Trenkwalder}}, \bibinfo
  {author} {\bibfnamefont {G.}~\bibnamefont {Semeghini}}, \bibinfo {author}
  {\bibfnamefont {G.}~\bibnamefont {Spagnolli}}, \bibinfo {author}
  {\bibfnamefont {A.}~\bibnamefont {Simoni}}, \bibinfo {author} {\bibfnamefont
  {M.}~\bibnamefont {Fattori}}, \bibinfo {author} {\bibfnamefont
  {M.}~\bibnamefont {Inguscio}}, \ and\ \bibinfo {author} {\bibfnamefont
  {G.}~\bibnamefont {Modugno}},\ }\href {\doibase
  10.1103/PhysRevLett.111.053202} {\bibfield  {journal} {\bibinfo  {journal}
  {Phys. Rev. Lett.}\ }\textbf {\bibinfo {volume} {111}},\ \bibinfo {pages}
  {053202} (\bibinfo {year} {2013})}\BibitemShut {NoStop}%
\bibitem [{\citenamefont {Wild}\ \emph {et~al.}(2012)\citenamefont {Wild},
  \citenamefont {Makotyn}, \citenamefont {Pino}, \citenamefont {Cornell},\ and\
  \citenamefont {Jin}}]{Wild2012}%
  \BibitemOpen
  \bibfield  {author} {\bibinfo {author} {\bibfnamefont {R.~J.}\ \bibnamefont
  {Wild}}, \bibinfo {author} {\bibfnamefont {P.}~\bibnamefont {Makotyn}},
  \bibinfo {author} {\bibfnamefont {J.~M.}\ \bibnamefont {Pino}}, \bibinfo
  {author} {\bibfnamefont {E.~A.}\ \bibnamefont {Cornell}}, \ and\ \bibinfo
  {author} {\bibfnamefont {D.~S.}\ \bibnamefont {Jin}},\ }\href {\doibase
  10.1103/PhysRevLett.108.145305} {\bibfield  {journal} {\bibinfo  {journal}
  {Phys. Rev. Lett.}\ }\textbf {\bibinfo {volume} {108}},\ \bibinfo {pages}
  {145305} (\bibinfo {year} {2012})}\BibitemShut {NoStop}%
\bibitem [{\citenamefont {Gross}\ \emph {et~al.}(2009)\citenamefont {Gross},
  \citenamefont {Shotan}, \citenamefont {Kokkelmans},\ and\ \citenamefont
  {Khaykovich}}]{Gross2009}%
  \BibitemOpen
  \bibfield  {author} {\bibinfo {author} {\bibfnamefont {N.}~\bibnamefont
  {Gross}}, \bibinfo {author} {\bibfnamefont {Z.}~\bibnamefont {Shotan}},
  \bibinfo {author} {\bibfnamefont {S.}~\bibnamefont {Kokkelmans}}, \ and\
  \bibinfo {author} {\bibfnamefont {L.}~\bibnamefont {Khaykovich}},\ }\href
  {\doibase 10.1103/PhysRevLett.103.163202} {\bibfield  {journal} {\bibinfo
  {journal} {Phys. Rev. Lett.}\ }\textbf {\bibinfo {volume} {103}},\ \bibinfo
  {pages} {163202} (\bibinfo {year} {2009})}\BibitemShut {NoStop}%
\bibitem [{\citenamefont {Gross}\ \emph {et~al.}(2010)\citenamefont {Gross},
  \citenamefont {Shotan}, \citenamefont {Kokkelmans},\ and\ \citenamefont
  {Khaykovich}}]{Gross2010}%
  \BibitemOpen
  \bibfield  {author} {\bibinfo {author} {\bibfnamefont {N.}~\bibnamefont
  {Gross}}, \bibinfo {author} {\bibfnamefont {Z.}~\bibnamefont {Shotan}},
  \bibinfo {author} {\bibfnamefont {S.}~\bibnamefont {Kokkelmans}}, \ and\
  \bibinfo {author} {\bibfnamefont {L.}~\bibnamefont {Khaykovich}},\ }\href
  {\doibase 10.1103/PhysRevLett.105.103203} {\bibfield  {journal} {\bibinfo
  {journal} {Phys. Rev. Lett.}\ }\textbf {\bibinfo {volume} {105}},\ \bibinfo
  {pages} {103203} (\bibinfo {year} {2010})}\BibitemShut {NoStop}%
\bibitem [{\citenamefont {Huang}\ \emph
  {et~al.}(2014{\natexlab{b}})\citenamefont {Huang}, \citenamefont {O'Hara},
  \citenamefont {Grimm}, \citenamefont {Hutson},\ and\ \citenamefont
  {Petrov}}]{Huang2014a}%
  \BibitemOpen
  \bibfield  {author} {\bibinfo {author} {\bibfnamefont {B.}~\bibnamefont
  {Huang}}, \bibinfo {author} {\bibfnamefont {K.~M.}\ \bibnamefont {O'Hara}},
  \bibinfo {author} {\bibfnamefont {R.}~\bibnamefont {Grimm}}, \bibinfo
  {author} {\bibfnamefont {J.~M.}\ \bibnamefont {Hutson}}, \ and\ \bibinfo
  {author} {\bibfnamefont {D.~S.}\ \bibnamefont {Petrov}},\ }\href {\doibase
  10.1103/PhysRevA.90.043636} {\bibfield  {journal} {\bibinfo  {journal} {Phys.
  Rev. A}\ }\textbf {\bibinfo {volume} {90}},\ \bibinfo {pages} {043636}
  (\bibinfo {year} {2014}{\natexlab{b}})}\BibitemShut {NoStop}%
\bibitem [{\citenamefont {Huang}\ \emph {et~al.}(2015)\citenamefont {Huang},
  \citenamefont {Sidorenkov},\ and\ \citenamefont {Grimm}}]{Huang2015a}%
  \BibitemOpen
  \bibfield  {author} {\bibinfo {author} {\bibfnamefont {B.}~\bibnamefont
  {Huang}}, \bibinfo {author} {\bibfnamefont {L.~A.}\ \bibnamefont
  {Sidorenkov}}, \ and\ \bibinfo {author} {\bibfnamefont {R.}~\bibnamefont
  {Grimm}},\ }\href {\doibase 10.1103/PhysRevA.91.063622} {\bibfield  {journal}
  {\bibinfo  {journal} {Phys. Rev. A}\ }\textbf {\bibinfo {volume} {91}},\
  \bibinfo {pages} {063622} (\bibinfo {year} {2015})}\BibitemShut {NoStop}%
\bibitem [{\citenamefont {Wang}\ \emph
  {et~al.}(2012{\natexlab{b}})\citenamefont {Wang}, \citenamefont {Wang},
  \citenamefont {D'Incao},\ and\ \citenamefont {Greene}}]{Wang2012d}%
  \BibitemOpen
  \bibfield  {author} {\bibinfo {author} {\bibfnamefont {Y.}~\bibnamefont
  {Wang}}, \bibinfo {author} {\bibfnamefont {J.}~\bibnamefont {Wang}}, \bibinfo
  {author} {\bibfnamefont {J.~P.}\ \bibnamefont {D'Incao}}, \ and\ \bibinfo
  {author} {\bibfnamefont {C.~H.}\ \bibnamefont {Greene}},\ }\href {\doibase
  10.1103/PhysRevLett.109.243201} {\bibfield  {journal} {\bibinfo  {journal}
  {Phys. Rev. Lett.}\ }\textbf {\bibinfo {volume} {109}},\ \bibinfo {pages}
  {243201} (\bibinfo {year} {2012}{\natexlab{b}})}\BibitemShut {NoStop}%
\bibitem [{\citenamefont {Maier}\ \emph {et~al.}(2015)\citenamefont {Maier},
  \citenamefont {Eisele}, \citenamefont {Tiemann},\ and\ \citenamefont
  {Zimmermann}}]{Maier2015}%
  \BibitemOpen
  \bibfield  {author} {\bibinfo {author} {\bibfnamefont {R.~A.~W.}\
  \bibnamefont {Maier}}, \bibinfo {author} {\bibfnamefont {M.}~\bibnamefont
  {Eisele}}, \bibinfo {author} {\bibfnamefont {E.}~\bibnamefont {Tiemann}}, \
  and\ \bibinfo {author} {\bibfnamefont {C.}~\bibnamefont {Zimmermann}},\
  }\href {\doibase 10.1103/PhysRevLett.115.043201} {\bibfield  {journal}
  {\bibinfo  {journal} {Phys. Rev. Lett.}\ }\textbf {\bibinfo {volume} {115}},\
  \bibinfo {pages} {043201} (\bibinfo {year} {2015})}\BibitemShut {NoStop}%
\bibitem [{\citenamefont {Barontini}\ \emph {et~al.}(2009)\citenamefont
  {Barontini}, \citenamefont {Weber}, \citenamefont {Rabatti}, \citenamefont
  {Catani}, \citenamefont {Thalhammer}, \citenamefont {Inguscio},\ and\
  \citenamefont {Minardi}}]{Barontini2009}%
  \BibitemOpen
  \bibfield  {author} {\bibinfo {author} {\bibfnamefont {G.}~\bibnamefont
  {Barontini}}, \bibinfo {author} {\bibfnamefont {C.}~\bibnamefont {Weber}},
  \bibinfo {author} {\bibfnamefont {F.}~\bibnamefont {Rabatti}}, \bibinfo
  {author} {\bibfnamefont {J.}~\bibnamefont {Catani}}, \bibinfo {author}
  {\bibfnamefont {G.}~\bibnamefont {Thalhammer}}, \bibinfo {author}
  {\bibfnamefont {M.}~\bibnamefont {Inguscio}}, \ and\ \bibinfo {author}
  {\bibfnamefont {F.}~\bibnamefont {Minardi}},\ }\href {\doibase
  10.1103/PhysRevLett.103.043201} {\bibfield  {journal} {\bibinfo  {journal}
  {Phys. Rev. Lett.}\ }\textbf {\bibinfo {volume} {103}},\ \bibinfo {pages}
  {043201} (\bibinfo {year} {2009})}\BibitemShut {NoStop}%
\bibitem [{\citenamefont {Bloom}\ \emph {et~al.}(2013)\citenamefont {Bloom},
  \citenamefont {Hu}, \citenamefont {Cumby},\ and\ \citenamefont
  {Jin}}]{Bloom2013}%
  \BibitemOpen
  \bibfield  {author} {\bibinfo {author} {\bibfnamefont {R.~S.}\ \bibnamefont
  {Bloom}}, \bibinfo {author} {\bibfnamefont {M.-G.}\ \bibnamefont {Hu}},
  \bibinfo {author} {\bibfnamefont {T.~D.}\ \bibnamefont {Cumby}}, \ and\
  \bibinfo {author} {\bibfnamefont {D.~S.}\ \bibnamefont {Jin}},\ }\href
  {\doibase 10.1103/PhysRevLett.111.105301} {\bibfield  {journal} {\bibinfo
  {journal} {Phys. Rev. Lett.}\ }\textbf {\bibinfo {volume} {111}},\ \bibinfo
  {pages} {105301} (\bibinfo {year} {2013})}\BibitemShut {NoStop}%
\bibitem [{\citenamefont {{Wacker}}\ \emph {et~al.}(2016)\citenamefont
  {{Wacker}}, \citenamefont {{J{\o}rgensen}}, \citenamefont {{Birkmose}},
  \citenamefont {{Winter}}, \citenamefont {{Mikkelsen}}, \citenamefont
  {{Sherson}}, \citenamefont {{Zinner}},\ and\ \citenamefont
  {{Arlt}}}]{Wacker2016}%
  \BibitemOpen
  \bibfield  {author} {\bibinfo {author} {\bibfnamefont {L.~J.}\ \bibnamefont
  {{Wacker}}}, \bibinfo {author} {\bibfnamefont {N.~B.}\ \bibnamefont
  {{J{\o}rgensen}}}, \bibinfo {author} {\bibfnamefont {D.}~\bibnamefont
  {{Birkmose}}}, \bibinfo {author} {\bibfnamefont {N.}~\bibnamefont
  {{Winter}}}, \bibinfo {author} {\bibfnamefont {M.}~\bibnamefont
  {{Mikkelsen}}}, \bibinfo {author} {\bibfnamefont {J.}~\bibnamefont
  {{Sherson}}}, \bibinfo {author} {\bibfnamefont {N.}~\bibnamefont {{Zinner}}},
  \ and\ \bibinfo {author} {\bibfnamefont {J.~J.}\ \bibnamefont {{Arlt}}},\
  }\href@noop {} {\bibfield  {journal} {\bibinfo  {journal} {ArXiv e-prints}\ }
  (\bibinfo {year} {2016})},\ \Eprint {http://arxiv.org/abs/1604.03693}
  {arXiv:1604.03693 [cond-mat.quant-gas]} \BibitemShut {NoStop}%
\bibitem [{\citenamefont {Repp}\ \emph {et~al.}(2013)\citenamefont {Repp},
  \citenamefont {Pires}, \citenamefont {Ulmanis}, \citenamefont {Heck},
  \citenamefont {Kuhnle}, \citenamefont {Weidem\"uller},\ and\ \citenamefont
  {Tiemann}}]{Repp2013}%
  \BibitemOpen
  \bibfield  {author} {\bibinfo {author} {\bibfnamefont {M.}~\bibnamefont
  {Repp}}, \bibinfo {author} {\bibfnamefont {R.}~\bibnamefont {Pires}},
  \bibinfo {author} {\bibfnamefont {J.}~\bibnamefont {Ulmanis}}, \bibinfo
  {author} {\bibfnamefont {R.}~\bibnamefont {Heck}}, \bibinfo {author}
  {\bibfnamefont {E.~D.}\ \bibnamefont {Kuhnle}}, \bibinfo {author}
  {\bibfnamefont {M.}~\bibnamefont {Weidem\"uller}}, \ and\ \bibinfo {author}
  {\bibfnamefont {E.}~\bibnamefont {Tiemann}},\ }\href {\doibase
  10.1103/PhysRevA.87.010701} {\bibfield  {journal} {\bibinfo  {journal} {Phys.
  Rev. A}\ }\textbf {\bibinfo {volume} {87}},\ \bibinfo {pages} {010701}
  (\bibinfo {year} {2013})}\BibitemShut {NoStop}%
\bibitem [{\citenamefont {Tung}\ \emph {et~al.}(2013)\citenamefont {Tung},
  \citenamefont {Parker}, \citenamefont {Johansen}, \citenamefont {Chin},
  \citenamefont {Wang},\ and\ \citenamefont {Julienne}}]{Tung2013}%
  \BibitemOpen
  \bibfield  {author} {\bibinfo {author} {\bibfnamefont {S.-K.}\ \bibnamefont
  {Tung}}, \bibinfo {author} {\bibfnamefont {C.}~\bibnamefont {Parker}},
  \bibinfo {author} {\bibfnamefont {J.}~\bibnamefont {Johansen}}, \bibinfo
  {author} {\bibfnamefont {C.}~\bibnamefont {Chin}}, \bibinfo {author}
  {\bibfnamefont {Y.}~\bibnamefont {Wang}}, \ and\ \bibinfo {author}
  {\bibfnamefont {P.~S.}\ \bibnamefont {Julienne}},\ }\href {\doibase
  10.1103/PhysRevA.87.010702} {\bibfield  {journal} {\bibinfo  {journal} {Phys.
  Rev. A}\ }\textbf {\bibinfo {volume} {87}},\ \bibinfo {pages} {010702}
  (\bibinfo {year} {2013})}\BibitemShut {NoStop}%
\bibitem [{\citenamefont {Pires}\ \emph
  {et~al.}(2014{\natexlab{b}})\citenamefont {Pires}, \citenamefont {Repp},
  \citenamefont {Ulmanis}, \citenamefont {Kuhnle}, \citenamefont
  {Weidem\"uller}, \citenamefont {Tiecke}, \citenamefont {Greene},
  \citenamefont {Ruzic}, \citenamefont {Bohn},\ and\ \citenamefont
  {Tiemann}}]{Pires2014a}%
  \BibitemOpen
  \bibfield  {author} {\bibinfo {author} {\bibfnamefont {R.}~\bibnamefont
  {Pires}}, \bibinfo {author} {\bibfnamefont {M.}~\bibnamefont {Repp}},
  \bibinfo {author} {\bibfnamefont {J.}~\bibnamefont {Ulmanis}}, \bibinfo
  {author} {\bibfnamefont {E.~D.}\ \bibnamefont {Kuhnle}}, \bibinfo {author}
  {\bibfnamefont {M.}~\bibnamefont {Weidem\"uller}}, \bibinfo {author}
  {\bibfnamefont {T.~G.}\ \bibnamefont {Tiecke}}, \bibinfo {author}
  {\bibfnamefont {C.~H.}\ \bibnamefont {Greene}}, \bibinfo {author}
  {\bibfnamefont {B.~P.}\ \bibnamefont {Ruzic}}, \bibinfo {author}
  {\bibfnamefont {J.~L.}\ \bibnamefont {Bohn}}, \ and\ \bibinfo {author}
  {\bibfnamefont {E.}~\bibnamefont {Tiemann}},\ }\href {\doibase
  10.1103/PhysRevA.90.012710} {\bibfield  {journal} {\bibinfo  {journal} {Phys.
  Rev. A}\ }\textbf {\bibinfo {volume} {90}},\ \bibinfo {pages} {012710}
  (\bibinfo {year} {2014}{\natexlab{b}})}\BibitemShut {NoStop}%
\bibitem [{Note1()}]{Note1}%
  \BibitemOpen
  \bibinfo {note} {In fact a Landau-Zener estimate shows that for $a_{BX}$ near
  unitarity, these potentials are midway between the diabatic and adiabatic
  limits for Cs-Cs-Li, but for $|a_{BX}|<10|a_{BB}|$ they are closer to the
  adiabatic limit.}\BibitemShut {Stop}%
\bibitem [{Note2()}]{Note2}%
  \BibitemOpen
  \bibinfo {note} {The Li-Cs scattering length is inferred from the
  radio-frequency association of LiCs Feshbach dimers around 843~G and 889~G
  and a coupled channels calculation~\cite {Ulmanis2015}. The corresponding
  Cs-Cs scattering lengths are obtained using the parametrization given in
  Ref.~\cite {Berninger2013}.}\BibitemShut {Stop}%
\bibitem [{\citenamefont {Berninger}\ \emph {et~al.}(2013)\citenamefont
  {Berninger}, \citenamefont {Zenesini}, \citenamefont {Huang}, \citenamefont
  {Harm}, \citenamefont {N\"agerl}, \citenamefont {Ferlaino}, \citenamefont
  {Grimm}, \citenamefont {Julienne},\ and\ \citenamefont
  {Hutson}}]{Berninger2013}%
  \BibitemOpen
  \bibfield  {author} {\bibinfo {author} {\bibfnamefont {M.}~\bibnamefont
  {Berninger}}, \bibinfo {author} {\bibfnamefont {A.}~\bibnamefont {Zenesini}},
  \bibinfo {author} {\bibfnamefont {B.}~\bibnamefont {Huang}}, \bibinfo
  {author} {\bibfnamefont {W.}~\bibnamefont {Harm}}, \bibinfo {author}
  {\bibfnamefont {H.-C.}\ \bibnamefont {N\"agerl}}, \bibinfo {author}
  {\bibfnamefont {F.}~\bibnamefont {Ferlaino}}, \bibinfo {author}
  {\bibfnamefont {R.}~\bibnamefont {Grimm}}, \bibinfo {author} {\bibfnamefont
  {P.~S.}\ \bibnamefont {Julienne}}, \ and\ \bibinfo {author} {\bibfnamefont
  {J.~M.}\ \bibnamefont {Hutson}},\ }\href {\doibase
  10.1103/PhysRevA.87.032517} {\bibfield  {journal} {\bibinfo  {journal} {Phys.
  Rev. A}\ }\textbf {\bibinfo {volume} {87}},\ \bibinfo {pages} {032517}
  (\bibinfo {year} {2013})}\BibitemShut {NoStop}%
\bibitem [{Note3()}]{Note3}%
  \BibitemOpen
  \bibinfo {note} {Note that this definition of $ L_3 $ differs by a factor of
  two from the one that was used in Ref.~\cite {Pires2014}.}\BibitemShut
  {Stop}%
\bibitem [{\citenamefont {Weber}\ \emph {et~al.}(2003)\citenamefont {Weber},
  \citenamefont {Herbig}, \citenamefont {Mark}, \citenamefont {N\"agerl},\ and\
  \citenamefont {Grimm}}]{Weber2003}%
  \BibitemOpen
  \bibfield  {author} {\bibinfo {author} {\bibfnamefont {T.}~\bibnamefont
  {Weber}}, \bibinfo {author} {\bibfnamefont {J.}~\bibnamefont {Herbig}},
  \bibinfo {author} {\bibfnamefont {M.}~\bibnamefont {Mark}}, \bibinfo {author}
  {\bibfnamefont {H.-C.}\ \bibnamefont {N\"agerl}}, \ and\ \bibinfo {author}
  {\bibfnamefont {R.}~\bibnamefont {Grimm}},\ }\href {\doibase
  10.1103/PhysRevLett.91.123201} {\bibfield  {journal} {\bibinfo  {journal}
  {Phys. Rev. Lett.}\ }\textbf {\bibinfo {volume} {91}},\ \bibinfo {pages}
  {123201} (\bibinfo {year} {2003})}\BibitemShut {NoStop}%
\bibitem [{\citenamefont {D'Incao}\ and\ \citenamefont
  {Esry}(2009)}]{DIncao2009a}%
  \BibitemOpen
  \bibfield  {author} {\bibinfo {author} {\bibfnamefont {J.~P.}\ \bibnamefont
  {D'Incao}}\ and\ \bibinfo {author} {\bibfnamefont {B.~D.}\ \bibnamefont
  {Esry}},\ }\href {\doibase 10.1103/PhysRevLett.103.083202} {\bibfield
  {journal} {\bibinfo  {journal} {Phys. Rev. Lett.}\ }\textbf {\bibinfo
  {volume} {103}},\ \bibinfo {pages} {083202} (\bibinfo {year}
  {2009})}\BibitemShut {NoStop}%
\bibitem [{tbp()}]{tbp}%
  \BibitemOpen
  \href@noop {} {\ }\bibinfo {note} {To be published}\BibitemShut {NoStop}%
\bibitem [{\citenamefont {S\o{}rensen}\ \emph {et~al.}(2012)\citenamefont
  {S\o{}rensen}, \citenamefont {Fedorov}, \citenamefont {Jensen},\ and\
  \citenamefont {Zinner}}]{Sorensen2012}%
  \BibitemOpen
  \bibfield  {author} {\bibinfo {author} {\bibfnamefont {P.~K.}\ \bibnamefont
  {S\o{}rensen}}, \bibinfo {author} {\bibfnamefont {D.~V.}\ \bibnamefont
  {Fedorov}}, \bibinfo {author} {\bibfnamefont {A.~S.}\ \bibnamefont {Jensen}},
  \ and\ \bibinfo {author} {\bibfnamefont {N.~T.}\ \bibnamefont {Zinner}},\
  }\href {\doibase 10.1103/PhysRevA.86.052516} {\bibfield  {journal} {\bibinfo
  {journal} {Phys. Rev. A}\ }\textbf {\bibinfo {volume} {86}},\ \bibinfo
  {pages} {052516} (\bibinfo {year} {2012})}\BibitemShut {NoStop}%
\bibitem [{\citenamefont {Petrov}\ and\ \citenamefont
  {Werner}(2015)}]{Petrov2015}%
  \BibitemOpen
  \bibfield  {author} {\bibinfo {author} {\bibfnamefont {D.~S.}\ \bibnamefont
  {Petrov}}\ and\ \bibinfo {author} {\bibfnamefont {F.}~\bibnamefont
  {Werner}},\ }\href {\doibase 10.1103/PhysRevA.92.022704} {\bibfield
  {journal} {\bibinfo  {journal} {Phys. Rev. A}\ }\textbf {\bibinfo {volume}
  {92}},\ \bibinfo {pages} {022704} (\bibinfo {year} {2015})}\BibitemShut
  {NoStop}%
\bibitem [{\citenamefont {{Mikkelsen}}\ \emph {et~al.}(2015)\citenamefont
  {{Mikkelsen}}, \citenamefont {{Jensen}}, \citenamefont {{Fedorov}},\ and\
  \citenamefont {{Zinner}}}]{Mikkelsen2015}%
  \BibitemOpen
  \bibfield  {author} {\bibinfo {author} {\bibfnamefont {M.}~\bibnamefont
  {{Mikkelsen}}}, \bibinfo {author} {\bibfnamefont {A.~S.}\ \bibnamefont
  {{Jensen}}}, \bibinfo {author} {\bibfnamefont {D.~V.}\ \bibnamefont
  {{Fedorov}}}, \ and\ \bibinfo {author} {\bibfnamefont {N.~T.}\ \bibnamefont
  {{Zinner}}},\ }\href {\doibase 10.1088/0953-4075/48/8/085301} {\bibfield
  {journal} {\bibinfo  {journal} {J. Phys. B: At., Mol. Opt. Phys.}\ }\textbf
  {\bibinfo {volume} {48}},\ \bibinfo {pages} {085301} (\bibinfo {year}
  {2015})}\BibitemShut {NoStop}%
\bibitem [{\citenamefont {D'Incao}\ and\ \citenamefont
  {Esry}(2006)}]{DIncao2006}%
  \BibitemOpen
  \bibfield  {author} {\bibinfo {author} {\bibfnamefont {J.~P.}\ \bibnamefont
  {D'Incao}}\ and\ \bibinfo {author} {\bibfnamefont {B.~D.}\ \bibnamefont
  {Esry}},\ }\href {\doibase 10.1103/PhysRevA.73.030702} {\bibfield  {journal}
  {\bibinfo  {journal} {Phys. Rev. A}\ }\textbf {\bibinfo {volume} {73}},\
  \bibinfo {pages} {030702} (\bibinfo {year} {2006})}\BibitemShut {NoStop}%
\bibitem [{\citenamefont {Helfrich}\ and\ \citenamefont
  {Hammer}(2010)}]{Helfrich2010a}%
  \BibitemOpen
  \bibfield  {author} {\bibinfo {author} {\bibfnamefont {K.}~\bibnamefont
  {Helfrich}}\ and\ \bibinfo {author} {\bibfnamefont {H.-W.}\ \bibnamefont
  {Hammer}},\ }\href {\doibase 10.1051/epjconf/20100302007} {\bibfield
  {journal} {\bibinfo  {journal} {EPJ Web}\ } (\bibinfo {year} {2010}),\
  10.1051/epjconf/20100302007}\BibitemShut {NoStop}%
\bibitem [{\citenamefont {Helfrich}\ \emph {et~al.}(2010)\citenamefont
  {Helfrich}, \citenamefont {Hammer},\ and\ \citenamefont
  {Petrov}}]{Helfrich2010}%
  \BibitemOpen
  \bibfield  {author} {\bibinfo {author} {\bibfnamefont {K.}~\bibnamefont
  {Helfrich}}, \bibinfo {author} {\bibfnamefont {H.-W.}\ \bibnamefont
  {Hammer}}, \ and\ \bibinfo {author} {\bibfnamefont {D.~S.}\ \bibnamefont
  {Petrov}},\ }\href {\doibase 10.1103/PhysRevA.81.042715} {\bibfield
  {journal} {\bibinfo  {journal} {Phys. Rev. A}\ }\textbf {\bibinfo {volume}
  {81}},\ \bibinfo {pages} {042715} (\bibinfo {year} {2010})}\BibitemShut
  {NoStop}%
\bibitem [{\citenamefont {Blume}\ and\ \citenamefont {Yan}(2014)}]{Blume2014}%
  \BibitemOpen
  \bibfield  {author} {\bibinfo {author} {\bibfnamefont {D.}~\bibnamefont
  {Blume}}\ and\ \bibinfo {author} {\bibfnamefont {Y.}~\bibnamefont {Yan}},\
  }\href {\doibase 10.1103/PhysRevLett.113.213201} {\bibfield  {journal}
  {\bibinfo  {journal} {Phys. Rev. Lett.}\ }\textbf {\bibinfo {volume} {113}},\
  \bibinfo {pages} {213201} (\bibinfo {year} {2014})}\BibitemShut {NoStop}%
\end{thebibliography}%

\end{document}